\documentclass[12pt]{article}
\usepackage[dvips]{graphicx,color}
\usepackage{amsmath}
\usepackage{amsfonts}
\begin{document}
\title{Chiral zero modes in non local domain walls}
\author{C.~D.~Fosco and ~G.~Torroba
  \\
  {\normalsize\it Centro At\'omico Bariloche and Instituto Balseiro}\\
  {\normalsize\it Comisi\'on Nacional de Energ\'\i a At\'omica}\\
  {\normalsize\it 8400 Bariloche, Argentina.}}  
\date{\today}
\maketitle
\begin{abstract}
\noindent
We study a generalization of the Callan-Harvey mechanism to the case
of a non local mass.  Using a $2+1$ model as a concrete example, we
show that both the existence and properties of localized zero modes
can also be consistently studied when the mass is non local.  After
dealing with some general properties of the resulting integral
equations, we show how non local masses naturally arise when radiative
corrections are included. We do that for a $2+1$ dimensional
example, and also evaluate the zero mode of the resulting non local
Dirac operator.  
\end{abstract}
\bigskip
\newpage


\section{Introduction}\label{intro}
The Callan-Harvey mechanism~\cite{Callan} explains the existence and
properties of the fermionic zero modes that appear whenever the mass
of a Dirac field in $2k+1$ dimensions ($k=1,2,\cdots$) has a domain-wall
like defect. The zero modes due to this phenomenon are localized,
concentrated around the domain wall, and chiral from the point of view
of the domain-wall world-volume (a $2k$-dimensional theory).

This mechanism has found many interesting applications, both to
phenomenological issues~\cite{condensed} and theoretical
elaborations~\cite{fl,fls}.  Dynamical domain walls have been
considered in~\cite{fradkin} and the very interesting case of a
dynamical supersymmetric model with this kind of configuration has
been discussed in~\cite{SUSY}.  Remarkably, lattice versions of the
zero modes, the so-called `domain wall fermions'~\cite{kap} have also
been precursors of the overlap Dirac operator~\cite{nar}, a sensible
way to put chiral fermions on a spacetime lattice.

In spite of the fact that the Callan-Harvey mechanism has been
extended in many directions, the existence of chiral zero modes has
been so far only studied for the case of local mass terms, namely,
those where the mass is a local function of the spacetime coordinates.

In this article, we consider a different generalization of the
problem, namely, the case of a non local mass in $2+1$ dimensions.
This is a non trivial modification of the usual assumptions of the
Callan-Harvey mechanism, which does, however, arise naturally in some
applications: we show that explicitly for a case where the fields are
dynamical, and radiative corrections are included.

The structure of the article is as follows: in section
\ref{sec:generalization} we set up the general framework, defining the
class of systems we shall consider, in terms of an associated
non-homogeneous integral equation for the zero mode. In section
\ref{sec:physmodel}, a concrete realization of the kind of non local
mass discussed earlier is exhibited, evaluating its zero mode
solutions. Section \ref{sec:conc} contains our conclusions.

\section{Callan-Harvey mechanism for a non local
  mass}\label{sec:generalization} 

In the standard Callan-Harvey mechanism~\cite{Callan} one considers a
fermion field in 2+1 dimensions coupled to a domain wall like defect,
with the Euclidean action

\begin{equation}
  \label{eq:action1}
S({\bar\psi},\psi) \;=\; \int d^3x \,{\bar\psi}\,[\not\!\partial
\,+\, m (x) ] \,\psi \;.
\end{equation} 
We use Euclidean coordinates  $x=(x_0 , x_1 , x_2)$ ($x_0$ is the Euclidean
time) and
$\not\!\partial= \gamma_\mu \partial_\mu\, ,$ where the $\gamma$-matrices are chosen
according to the convention:

\begin{equation}\label{dfgamma}
\gamma_0\;=\;\sigma_3 \;\;\;\gamma_1\;=\;\sigma_1 
\;\;\;\gamma_2\;=\;\sigma_2\;.
\end{equation}
The \textit{local} mass $m(x)$ contains a topological defect; in the
simplest case of a rectilinear static defect~\cite{fl}, they have the
characteristic shape:
\begin{equation}\label{eq:aux}
m(x) \sim \Lambda \; \sigma(x_2) \; ,
\end{equation}
where $\sigma(x_2) \equiv \rm{sign} (x_2)$.  Therefore the domain wall, which is
the interface between two regions with different signs for $m(x)$, is
the $x_1$ axis.

From the Dirac operator
\begin{equation}
  \label{eq:defDirac1}
{\mathcal D}\;=\; \not \! \partial _x + m(x_2) \;\;,
\end{equation}
we can construct the hermitian operator $\mathcal H  = {\mathcal D}^{\dag} \,
\mathcal D$. The form of $\mathcal H$ suggests the introduction of the adjoint
operators
\begin{equation}
  \label{eq:defa1}
a\;=\; \partial_2+m(x_2) \;\;,\;\;\;
a^ \dag\;=\;-\partial_2+m(x_2)
\end{equation}
in terms of which
\begin{equation}
\mathcal H \,=\, (a^ \dag a -{\not \! {\hat \partial}}^2)P_L+(a a^ \dag-{\not \! {\hat \partial}}^2)P_R 
\end{equation}
and
\begin{equation}
  \label{eq:Dirac1.1}
\mathcal D \,=\, (a+\not \! {\hat \partial})P_L+(a^ \dag+\not \! {\hat \partial})P_R 
\end{equation}
where $P_L= \frac{1}{2} (1+\gamma_2)$, $P_R= \frac{1}{2} (1-\gamma_2)$.
Expanding $\psi(x)$ in the complete set of eigenfunctions of $a^ \dag a$ and
$a a^ \dag$, there appears~\cite{fl} a massless left fermion, localized
over the domain wall. Its $x_2$ dependence is dictated by the fact
that it is a zero mode of the $a$ operator, and it dominates the low
energy dynamics of the system.

We want to generalize this phenomenon to include a non local mass (in
the following section we present a model that motivates this
generalization). Namely, rather than (\ref{eq:defDirac1}), the Dirac
operator shall be
\begin{equation}
  \label{eq:defDirac2}
\tilde{\mathcal D}(x,y)\;=\; \not \! \partial _x  \; \delta (x-y)+ M(x,y) \; \; .
\end{equation}
Little can be said about the existence of a fermionic zero mode before
we make some hypotheses to restrict the form of $M(x,y)$. We assume
that the system has translation invariance in the coordinates $\hat{x}
\equiv (x_0,x_1)$ and that $M(x,y)$ consists of a local domain wall like
part, plus a non local term with a strength controlled by a
parameter $\lambda$:
\begin{equation}
  \label{eq:defnonlocalmass}
M(x,y)\;=\; m(x_2) \; \delta (x-y)- \lambda \; \int\frac{d^2{\hat k}}{(2\pi)^2} \,
e^{i {\hat k} \cdot ({\hat x} - {\hat y})} \gamma_k (x_2,y_2)  \;\;.
\end{equation}

We are looking for a zero mode $\Psi(x)$, so that:
\begin{equation}
  \label{eq:defzeromode1}
\langle x |\tilde{\mathcal D}|\Psi \rangle\;=\; \int \; d^3y \; \tilde{\mathcal{D}} (x,y)
\Psi(y) =0 \; \; .
\end{equation}
Taking advantage of the translation invariance in $\hat x$, we use
`separation of variables' to look for solutions of the form:
\begin{equation}
  \label{eq:defPsi1}
\Psi(x)\;=\; \chi(\hat x) \, \psi(x_2) \; \; ,
\end{equation}
where $\chi(\hat x)$ is a massless spinor, which is
left-handed from the point of view of the two-dimensional world
defined by ${\hat x}$, i.e.,
\begin{equation}\label{eq:propchi}
\not \! {\hat \partial}  \chi(\hat x) \;=\; 0 \;\;\;
P_R  \chi(\hat x) \;=\; 0\;.
\end{equation}
There is an essential difference regarding the space of solutions to
the equations above in Euclidean and Minkowski spacetimes. Indeed, in
Euclidean spacetime, it leads to analytic functions of $x_0+i x_1$,
while in the Minkowski case one has `left-mover' solutions.  Keeping
this distinction in mind, we continue working with the Euclidean
version.

Substituting (\ref{eq:defPsi1}) into (\ref{eq:defzeromode1}) and
comparing with (\ref{eq:Dirac1.1}), we arrive to a non local version
of the kernel for the annihilation operator
\begin{equation}
  \label{eq:nonlocala1}
a(x_2,y_2)\;=\;[\partial_2+m(x_2)] \, \delta(x_2-y_2)- \lambda \; \gamma_k
(x_2,y_2) \;\;.
\end{equation}
The zero mode $\psi(x_2)$, must then satisfy the equation
\begin{equation}
  \label{eq:zeromode2}
{\langle x |a|\psi \rangle\;}=\;[\partial_2+m(x_2)] \, \psi(x_2)- \lambda \;
\int_{-\infty} ^{+\infty}
dy_2 \, \gamma_k(x_2,y_2) \psi(y_2)=0 \; \; .
\end{equation}
Following the method of variation of parameters, we use the ansatz
\begin{equation}
  \label{eq:zeromode3}
\psi(x_2)\,=\, \psi_0(x_2) \varphi(x_2)
\end{equation}
where $\psi_0$ is the zero mode for the local part in
(\ref{eq:zeromode2}), which satisfies
\begin{equation}
[\partial_2+m(x_2)] \, \psi_0(x_2)\;=\;0 \;\;\;\;\; \psi_0(x_2)\,=\, N \,
exp[-\int^{x_2}_0 ds
\, m(s)] \;,
\end{equation}
and $N$ is a normalization constant. 

The function $\varphi(x_2)$, which modulates $\psi_0(x_2)$ must then satisfy
the equation
\begin{equation}
\partial_2 \varphi(x_2)-\lambda \, \int_{-\infty}^{+\infty}
dy_2 \, [\psi_0 (x_2)^{-1} \, \gamma_k(x_2,y_2) \, \psi_0 (y_2)] \; \varphi (y_2)=0
\;.
\end{equation}
By integrating over $x_2$, the previous equation can be written in a
more convenient form as an integral equation
\begin{equation}
  \label{eq:fredholm1}
\varphi(x_2)-\lambda\,\int_{-\infty} ^{+\infty} dy_2 \, 
\tilde{\gamma}_k(x_2,y_2) \varphi(y_2) = \varphi(0)
\end{equation}
where the kernel $\tilde{\gamma}_k$ is
\begin{equation}
  \label{eq:kernel1}
\tilde{\gamma}_k(x_2,y_2)\;=\; \int_0 ^{x_2} dz_2 \, \psi_0 (z_2)^{-1} \,
\gamma_k(z_2,y_2) \, \psi_0 (y_2)\;.
\end{equation}
This is a homogeneous integral equation, which can be conveniently
rewritten as an equivalent non-homogeneous set of equations. Indeed,
introducing linear operators, (\ref{eq:fredholm1}) can be rewritten as
follows:
\begin{equation}
  \label{eq:fredholm2}
(I-\lambda \, T) \varphi \;=\;c
\end{equation}
where
\begin{equation}
(I \varphi)(x_2) \equiv \varphi(x_2)\; , \;\;\; c \equiv \varphi(0) \;,
\end{equation}
\begin{equation}
  \label{eq:fredholm3}
(T \varphi)(x_2) \equiv \int dy_2 \; {\tilde \gamma}_k (x_2,y_2) \, \varphi(y_2) \;.
\end{equation}
We see that the problem can be discussed in terms of
(\ref{eq:fredholm2}), which is an inhomogeneous system of the Fredholm
type~\cite{fredholm}. The condition $c \equiv \varphi(0)$ has to be verified, of
course, after solving (\ref{eq:fredholm2}) for arbitrary $c$.

The reason for this procedure is that, had we used the original
homogeneous system, we should have had to introduce non compact
operators, and the theory for this kind of operator is substantianlly
poorer than for the compact case.

\subsection{Properties of the solutions}
We have shown that the physical restrictions imposed on the non local
mass lead us to integral equations of the Fredholm type. The existence
and properties of solutions of these kinds of equations are very well
known~\cite{fredholm}. In particular, the theorem of the Fredholm
alternative provides information that is useful to our purposes.

The Fredholm alternative~\cite{fredholm} states that if $A=I-\lambda \, T$,
where $T$ is a compact operator on a Hilbert space $H$, then the
following alternative holds:
\begin{itemize}
\item{either $A \varphi_0 =0$ has only the trivial solution, in which case $A
\varphi = c$ has a unique solution $\forall c \in H$}
\item{or $A \varphi_0 =0$ has $q$ linearly independent solutions $\varphi_i \, \in
    H$. Then $A^\dagger {\tilde \varphi}_0 =0 $ also has $q$ linearly independent 
solutions ${\tilde \varphi}_i \, \in    H$. In this case $A \varphi=c$ is solvable 
iff $(c,{\tilde \varphi}_i)=0 \;\; \forall i=1, \dots, q$.
    
In the second alternative, the general non-homogeneous solution is
\begin{equation}
\varphi=\varphi_p+\sum_{i=1}^q a_i \varphi_i
\end{equation}
where $\varphi_p$ is a particular solution and $a_i$ are arbitrary
constants.  }
\end{itemize}

Note that, when the first alternative holds true, it implies in
particular that the solution will be unique when $c$ is replaced by
$\varphi(0)$. Solutions corresponding to $c \neq \varphi(0)$ are not solutions of the
system: \mbox{$(I-\lambda \, T) \varphi = c$}, \mbox{$c \equiv \varphi(0)$}, equivalent to
the original homogeneous equation (\ref{eq:fredholm1}), and may,
therefore, be discarded.

Any true solution of the system will also verify a subsidiary
equation, obtained by setting $x_2 = 0$ in (\ref{eq:fredholm1}):
\begin{equation}
\int_{-\infty} ^{+\infty} dy_2 \,\tilde{\gamma}_k(0,y_2) \varphi(y_2) = 0
\end{equation}
for  any $\lambda \neq 0$. This equation  shall be  true whenever the equations
$(I-\lambda \, T) \varphi = c$, $c \equiv \varphi(0)$ are both true (since it is derived from
them), and any solution will automatically verify it.

In our case, we want to study the effect of the non local term, the
strength of which is controlled by the value of $\lambda$. Close enough to the
local mass case, $\lambda$ can be made arbitrarily small, so
$\lambda^{-1}$ is not an eigenvalue of $T$ and consequently $A \varphi=0$ has
only the trivial solution. Therefore, if $T$ is a Fredholm operator,
(\ref{eq:fredholm1}) will have a unique solution. The functional space $H$ to
which $\varphi(x)$ belongs is restricted by the condition
\begin{equation}\label{eq:hcond}
\int dx_2 \; \Big(\psi_0(x_2) \Big)^2 \Big(\varphi(x_2)\Big)^2 < \infty  \; ,
\end{equation}
because $\psi(x)$ itself has to be normalizable.  This becomes a Hilbert
space, and it contains the zero mode of the local operator 
($\varphi\,=\, const$), when the scalar product used in $H$ is the one defined by
(\ref{eq:hcond}), namely,
\begin{equation}\label{eq:defscalar}
( f, g ) \;\equiv\; \int dx_2 \,\Big(\psi_0(x_2) \Big)^2 [f(x_2)]^* \, g(x_2) \;.
\end{equation}

Of course, things may be different if $H$ were equipped with the
standard $L^2$ scalar product, since then a constant function would not have a
finite norm, although it would verify (\ref{eq:hcond}). We assume
 that the norm defined by (\ref{eq:defscalar}) is used in what
follows.

When all the Fredholm's hypotheses are satisfied for a general kernel
$\tilde{\gamma}_k(x_2,y_2)$, it is possible to find a perturbative
solution. Indeed, writing
\begin{equation}
\varphi(x_2)\;=\;\varphi(0)+\lambda\,\int_{-\infty} ^\infty dy_2 \, 
\tilde{\gamma}_k(x_2,y_2) \, \varphi(y_2) \;,
\end{equation}
and successively replacing $\varphi(y_2)$ by $\varphi(0)+\lambda\,\int_{-\infty} ^\infty
dy_2 \, 
\tilde{\gamma}_k(x_2,y_2) \varphi(y_2)$ on the right hand side, we obtain the Neumann
series~\cite{fredholm}
\begin{equation}
\varphi(0)+\lambda \, K_1 (x_2) \, \varphi(0)+\dots+\lambda^n \, K_n (x_2)\, \varphi(0)+\dots
\end{equation}
where
\begin{equation}
  \label{eq:kernelj}
K_j (x_2) = \int _{-\infty} ^\infty \, dy_2 \, \gamma^{(j)}(x_2,y_2)
\end{equation}
and
\begin{eqnarray}
\gamma^{(1)}(x_2,y_2)&=& \tilde{\gamma}_k (x_2,y_2)\;\;,\;\;
\gamma^{(2)}(x_2,y_2)\;=\; \int_{-\infty}^\infty \, dz_2 \; \tilde{\gamma}_k (x_2,z_2)
\tilde{\gamma}_k (z_2,y_2)\nonumber\\
& \ldots & \gamma^{(n+1)}(x_2,y_2)\;=\; \int_{-\infty}^\infty \, dz_2 \; \tilde{\gamma}_k (x_2,z_2)
\gamma^{(n)} (z_2,y_2) \;.
\end{eqnarray}
It can be shown~\cite{fredholm} that the Neumann series converges
uniformly to the solution $\varphi(x)$ when 
\begin{equation}
  \label{eq:radiusconv}
|\lambda|<\frac{1}{\mu} \; ,
\end{equation}
\begin{equation}
  \label{eq:mucuad}
\mu^2=\int_{-\infty}^\infty \, dx_2 \, dy_2 \; [\tilde{\gamma}_k (x_2,y_2)]^2 \; .
\end{equation}
So choosing $\lambda$ to satisfy (\ref{eq:radiusconv}), we are allowed to represent
$\varphi$ by the expansion
\begin{equation}
  \label{eq:neumann}
\varphi(x_2)=\varphi(0)+\lambda \, K_1 (x_2) \, \varphi(0)+\dots+\lambda^n \,
K_n (x_2)\, \varphi(0)+\dots
\end{equation}

In subsection~\ref{subsec:quantumcorr} we shall use another method to
find a solution, which is useful when the kernel is
separable~\cite{fredholm}. Since this condition may be satisfied by an
operator which is not of the Fredholm type, it allows us to explore
rather different situations.


\subsection{Example}
After the previous general analysis, we now consider the effects of
quantum corrections for a specific choice of both the local and non
local parts of $M(x,y)$ in (\ref{eq:defnonlocalmass}). 
A natural generalization of the purely local mass case is to have
\mbox{$m(x_2)= \Lambda \, \sigma(x_2)$} and a $\gamma_k(x_2,y_2)$ which is `strongly
diagonal' and symmetric in $(x_2,y_2)$, i.e.:
\begin{equation}
  \label{eq:kernelspecific}
\, \gamma_k(x_2,y_2) \,=\,\frac{1}{2} \,
\Big[\sigma(x_2)+\sigma(y_2)\Big]\delta_N (x_2-y_2) \;,
\end{equation}
where $\delta_N$ is an approximation of Dirac's delta: $\delta_N (x_2-y_2) \to
\delta(x_2-y_2)$ when $N\to\infty$. Note that $\gamma_k(x_2,y_2) \to {\rm sign}(x_2) \, \delta(x_2-y_2)$ when
$N\to\infty$, so that the non local term reduces to a local domain wall mass.

We then adopt `natural' units such that $\Lambda= 1$, and study the
particular case $|x_2|\leq L = \Lambda^{-1}$. Regarding $\delta_N$, we use a truncation of
the one-dimensional completeness relation:
\begin{equation}
\delta_N (x_2,y_2)\,=\,\sum_{n=0}^N \, \varphi_n(x_2) \, \varphi_n
^{\dag}(y_2) \;,
\end{equation}
where $\{\varphi_n \}$ is a complete set of functions. We chose $\varphi_n(x)$ to
be the harmonic oscillator's eigenfunctions. 

The normalized zero mode corresponding to the local part is
\begin{equation}
  \label{eq:zeromodelocal2}
\psi_0 (x_2) \,=\,N_0 \, \exp \Big[-\int_0 ^{x_2} dt \, m(t)
\Big]\,=\,\frac{1}{\sqrt{1-e^{-2}}} \, e^{-|x_2|} \; ,
\end{equation}
and the corrected zero mode $\psi(x_2)\,=\,\psi_0 (x_2) \, \varphi(x_2) \;$ is
then determined by $\varphi(x_2)$. The integral equation for $\varphi(x_2)$
becomes
\begin{equation}
  \label{eq:integrequat}
\varphi(x_2)\,=\,\varphi(0)+\lambda \, \int_{-1}^1 dy_2 \;
\tilde{\gamma}(x_2,y_2) \, \varphi(y_2)
\end{equation}
where
$$
\tilde{\gamma}(x_2,y_2) \,=\,\frac{1}{2} \int_0 ^{x_2} \,dz_2 \; e^{|z_2|}
\, \Big[\sigma (z_2) \, + \, \sigma (y_2) \Big] \,
$$
\begin{equation}
  \label{eq:gammacompact}
\times \Big[ \sum_{n=0}^N \, \varphi_n(z_2) \, \varphi_n^{\dag}(y_2) \Big ] \, e^{-|y_2|} \; .
\end{equation}

\begin{figure}
\begin{center}
\begin{picture}(0,0)%
\includegraphics{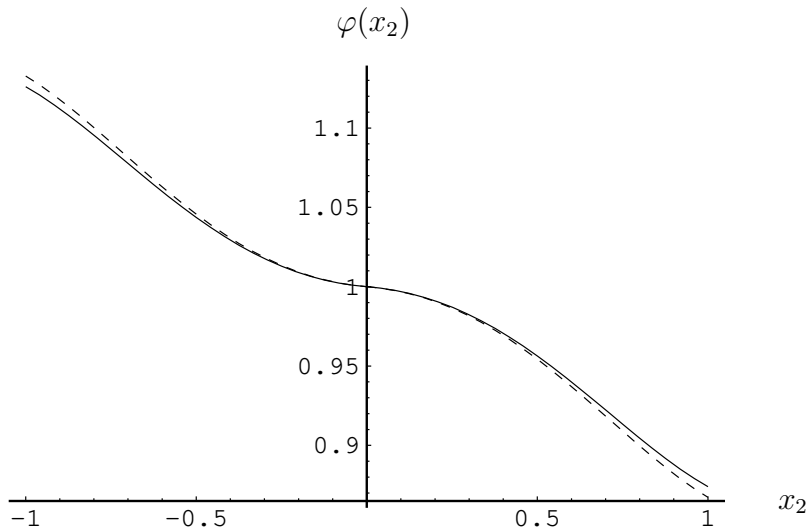}%
\end{picture}%
\setlength{\unitlength}{3947sp}%
\begingroup\makeatletter\ifx\SetFigFont\undefined%
\gdef\SetFigFont#1#2#3#4#5{%
  \reset@font\fontsize{#1}{#2pt}%
  \fontfamily{#3}\fontseries{#4}\fontshape{#5}%
  \selectfont}%
\fi\endgroup%
\begin{picture}(4855,3254)(1596,-12246)
\put(3676,-9136){\makebox(0,0)[lb]{\smash{\SetFigFont{12}{14.4}{\rmdefault}{\mddefault}{\updefault}{\color[rgb]{0,0,0}$\varphi(x_2)$}%
}}}
\put(6451,-12136){\makebox(0,0)[lb]{\smash{\SetFigFont{12}{14.4}{\rmdefault}{\mddefault}{\updefault}{\color[rgb]{0,0,0}$x_2$}%
}}}
\end{picture}
\end{center}
\caption{\footnotesize{$\varphi(x_2)\,$ after two iterations of the Neumann series
(\ref{eq:neumann}) for the $x_2$-compactified case. The dashed line corresponds
to the first iteration.}}
\label{varphicompact}
\end{figure}

From equations (\ref{eq:radiusconv}) and (\ref{eq:mucuad}), we see
that the Neumann series converges, in this case, for $|\lambda|<0.0990$. So
we assume $\lambda=0.01$, and we are ready to calculate
(\ref{eq:integrequat}) pertubatively. Since its expression in terms of
analytic functions is not very illuminating, we present, in
Figure~\ref{varphicompact}, the numerical results of the first two
iterations, taking $\varphi(0)=1\,$ and $N=3\,$.

\begin{figure}
\begin{center}
\begin{picture}(0,0)%
\includegraphics{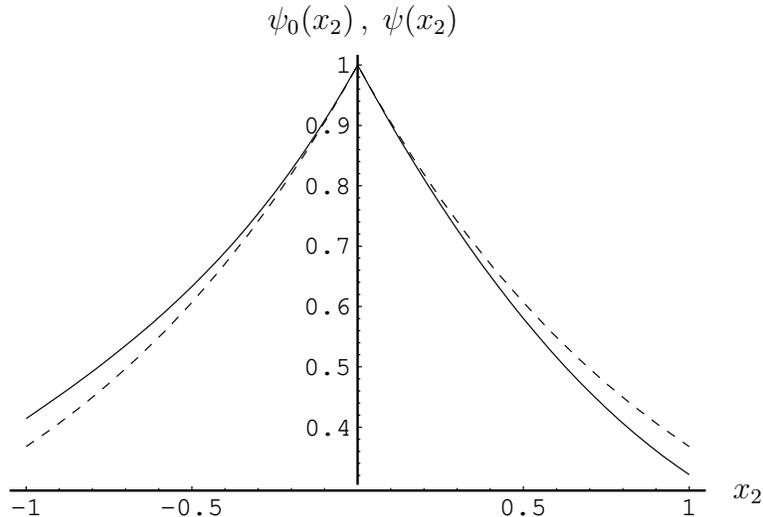}%
\end{picture}%
\setlength{\unitlength}{3947sp}%
\begingroup\makeatletter\ifx\SetFigFont\undefined%
\gdef\SetFigFont#1#2#3#4#5{%
  \reset@font\fontsize{#1}{#2pt}%
  \fontfamily{#3}\fontseries{#4}\fontshape{#5}%
  \selectfont}%
\fi\endgroup%
\begin{picture}(4567,3199)(1659,-12246)
\put(6226,-12136){\makebox(0,0)[lb]{\smash{\SetFigFont{12}{14.4}{\rmdefault}{\mddefault}{\updefault}{\color[rgb]{0,0,0}$x_2$}%
}}}
\put(3301,-9191){\makebox(0,0)[lb]{\smash{\SetFigFont{12}{14.4}{\rmdefault}{\mddefault}{\updefault}{\color[rgb]{0,0,0}$\psi_0 (x_2) \, , \; \psi(x_2)$}%
}}}
\end{picture}
\end{center}
\caption{\footnotesize{Zero mode profiles for  local (dashed line) and
    non local (full line) masses.}}
\label{compactmodes}
\end{figure}

In Figure~\ref{compactmodes}, we show the zero mode
(\ref{eq:zeromodelocal2}) and the one with $\lambda^2$-order corrections. We
see that the corrected zero mode continues to be localized over the
domain wall, although it is no longer a symmetric function of $x_2$.
Besides, this perturbative method introduces only smooth corrections
in the zero mode, as expected.

\section{A physical model for the non local mass}\label{sec:physmodel}
A natural way of generating a non local mass of the form
(\ref{eq:defnonlocalmass}) is to couple the Dirac field to a defect
that is allowed to fluctuate, with a dynamics given by a scalar field
action. In this context, quantum corrections to the classical
configurations generate a non local term in the Dirac operator,
already at the one loop order.  We shall now derive such a term, and
study how the classical fermionic zero mode is affected by its
presence.

\subsection{Description of the system}
We consider the Euclidean action
\begin{equation}\label{eq:euclaction}
S(\phi,{\bar\psi},\psi)\;=\; S_B (\phi) \,+\, S_F ({\bar\psi},\psi,\phi) 
\end{equation}
where
\begin{equation}
S_B(\phi) \;=\; \int d^3x \,[\frac{1}{2} (\partial \cdot \phi)^2  \,+\,
V(\phi) ] \;,
\end{equation}
\begin{equation}
S_F({\bar\psi},\psi,\phi) \;=\; \int d^3x \,{\bar\psi}\,[\not\!\partial
\,+\, g \phi (x) ] \,\psi \;
\end{equation}
and $V(\phi)$ is the quartic potential:
\begin{equation}
  \label{eq:quarticV}
V(\phi)=\frac{g^2}{2} \Big(\phi^2-\frac{\kappa^2}{g^2} \Big)^2 \; .
\end{equation}
The reason for choosing this scalar potential is that if we only consider the
scalar field, there exists a domain wall like solution (known as the kink
configuration) of the form~\cite{Solitones}
\begin{equation}
  \label{eq:kink1}
\phi(x_2)=\frac{\kappa}{g} \; \rm{tanh}(\kappa \, x_2) \;.
\end{equation}

Now, for the system (\ref{eq:euclaction}), the Euler-Lagrange equations of motion
are 
$$
[\not\!\partial + g \phi(x) ] \psi \;=\;0 
$$
\begin{equation}
  \label{eq:eqmov1}
- \partial^2 \phi (x) \,+\, V'(\phi(x)) \,+\, g {\bar\psi}(x) \psi(x) \;=\; 0
\;.
\end{equation}
It is immediate to verify that the configuration given by the domain wall
(\ref{eq:kink1}) and the chiral zero mode
\begin{equation}
  \label{eq:zeromodeclassic}
\psi(x)\,=\,P_L \, \chi(\hat x) \; exp  \Big( -\int_0 ^{x_2} \, dt \, g \,
\phi(t)\Big) \;,
\end{equation}
with $\chi$ as in (\ref{eq:propchi}), is a self-consistent solution to the coupled
system of equations (\ref{eq:eqmov1}).

Quantum corrections appear in the vacuum expectation values (VEVs) $\langle
\phi \rangle \equiv \langle \Omega | \phi | \Omega \rangle \; , \;$ $\langle \psi \rangle \equiv \langle \Omega | \psi | \Omega \rangle$, where $|\Omega \rangle$ is
the interacting vacuum. The most direct way of calculating these
expectation values is within the context of the effective action
$\Gamma(\varphi,{\bar\chi},\chi)$~\cite{zinn}. Here $\varphi$, ${\bar\chi}$ and $\chi$ denote the
so called `classical fields' corresponding to $\phi$, ${\bar\psi}$ and $\psi$,
respectively.


\subsection{One loop calculations}
It is useful to consider the loopwise expansion~\cite{zinn}
\begin{equation}
  \label{eq:oneloop}
\Gamma(\varphi,{\bar\chi},\chi) \;=\; S(\varphi,{\bar\chi},\chi)
\,+\, \Gamma_1(\varphi,{\bar\chi},\chi) \,+\,\ldots
\end{equation}
where $S$ is the classical action (\ref{eq:euclaction}) and $\Gamma_1$ contains
one-loop diagrams that introduce $\hbar$-order corrections in the classical tree
graphs. After a direct calculation we find
\begin{equation}
  \label{eq:gamma1.1}
\Gamma_1 \;=\; -{\rm Tr}\ln{\mathcal D}+\frac{1}{2} {\rm Tr} \ln 
\left[ \frac{\delta^2 S_B(\varphi)}{\delta\phi(x_1)
\delta\phi(x_2)} \,-\,2 \, g^2 \, {\bar\chi}(x_1) {\mathcal D}^{-1}(x_1,x_2) 
\chi(x_2) \right] \;
\end{equation}
where ${\mathcal D}\;=\; \not \! \partial + g \varphi \;\;$. For simplicity, we
will approximate the field $\varphi(x_2)$ by taking
$$
\varphi(x_2)=\frac{\kappa}{g} \; \rm{tanh}(\kappa \, x_2) \,.
$$
As a result, the logarithms of operators in (\ref{eq:gamma1.1}) can be obtained
by Fourier-transforming in $\hat x$ and expanding in adequate complete sets of
functions in $x_2$. As before, these are eigenvectors of $a^{\dag}a$ and $a
a^{\dag}$, with $a\,=\,\partial_2+g\varphi(x_2)$. We shall note them as
$\psi_n(x_2)$ and $\tilde{\psi}_n(x_2)$:
\begin{equation}
  \label{eq:eigenvectors1}
a^{\dag}a \,\psi_n(x_2) \,=\,\lambda_n^2 \psi_n(x_2) \;,\;\;\; a
a^{\dag} \, \tilde{\psi}_n(x_2) \,=\, \lambda_n^2 \tilde{\psi}_n(x_2) \;.
\end{equation}

Only $a$ has a normalizable zero mode
\begin{equation}
  \label{eq:azeromode}
a \psi_0 (x_2)\,=\,0 \;,\;\; \psi_0(x_2)\,=\,\sqrt{\frac{\kappa}{2}} \,
\rm{sech}(\kappa x_2) \, .
\end{equation}
The rest of the $a^{\dag}\, a$ and $a \, a^{\dag}$ spectra coincide, and consist
of a continuum separated by a finite gap $\kappa^2\,>\,0$ from the zero mode. We
also need to introduce the operators
\begin{equation}
  \label{eq:defb}
b \,=\, \partial_2 + 2 \kappa \, \rm{tanh} (\kappa x_2) \;\;\;,\;\;\;
b^{\dag} \,=\, -\partial_2 + 2 \kappa \, \rm{tanh} (\kappa x_2) \;,
\end{equation}
with eigenfunctions $\xi_n(x_2) \; , \; \tilde{\xi}_n (x_2)$:
\begin{equation}
  \label{eq:defb2}
b^{\dag} \,b \, \xi_n(x_2) \,=\, \mu_n^2 \, \xi_n(x_2)\;\;\;,\;\;\;
b \, b^{\dag} \, \tilde{\xi}_n (x_2) \,=\,\mu_n^2 \, \tilde{\xi}_n (x_2).
\end{equation}
In terms of these operators,
$$
\Gamma_1 \,=\, -\frac{1}{2}{\rm Tr}\ln\left[-\hat{\partial}^2 + a^{\dag}\,a \right]
-\frac{1}{2}{\rm Tr}\ln\left[-\hat{\partial}^2+a \, a^{\dag} \right]
+\frac{1}{2}{\rm Tr}\ln\left[-\hat{\partial}^2 + b^{\dag} \,b \right]
$$
\begin{equation}
  \label{eq:gamma1.3}
+\, \frac{1}{2} {\rm Tr} \ln \left[ \delta(x_1-x_2) 
- 2 g^2 \int_y \Delta_\varphi(x_1,y) {\bar \chi}(y)
{\mathcal D}^{-1}(y,x_2) \chi(x_2) \right] \;.
\end{equation}

\begin{figure}
\begin{center}
\begin{picture}(0,0)%
\includegraphics{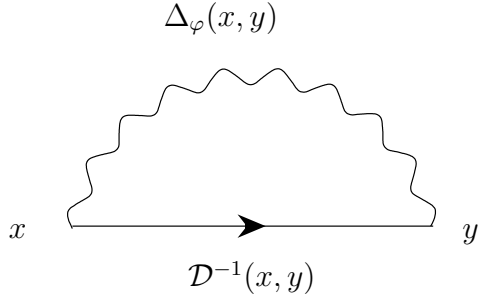}%
\end{picture}%
\setlength{\unitlength}{3947sp}%
\begingroup\makeatletter\ifx\SetFigFont\undefined%
\gdef\SetFigFont#1#2#3#4#5{%
  \reset@font\fontsize{#1}{#2pt}%
  \fontfamily{#3}\fontseries{#4}\fontshape{#5}%
  \selectfont}%
\fi\endgroup%
\begin{picture}(2850,1698)(1126,-3190)
\put(3976,-2836){\makebox(0,0)[lb]{\smash{\SetFigFont{12}{14.4}{\rmdefault}{\mddefault}{\updefault}{\color[rgb]{0,0,0}$y$}%
}}}
\put(1126,-2836){\makebox(0,0)[lb]{\smash{\SetFigFont{12}{14.4}{\rmdefault}{\mddefault}{\updefault}{\color[rgb]{0,0,0}$x$}%
}}}
\put(2251,-3136){\makebox(0,0)[lb]{\smash{\SetFigFont{12}{14.4}{\rmdefault}{\mddefault}{\updefault}{\color[rgb]{0,0,0}${\mathcal D}^{-1}(x,y)$}%
}}}
\put(2100,-1500){\makebox(0,0)[lb]{\smash{\SetFigFont{12}{14.4}{\rmdefault}{\mddefault}{\updefault}{\color[rgb]{0,0,0}$\Delta_{\varphi}(x,y)$}%
}}}
\end{picture}
\end{center}
\caption{\footnotesize{Feynman diagram for the one-loop quantum correction to
the Dirac operator.}}
\label{feynman}
\end{figure}

The quantum Dirac operator is then
\begin{equation}
  \label{eq:quantumD} 
{\tilde{\mathcal D}}(x,y)\;=\;-\frac{\delta^2\Gamma}{\delta{\bar\chi}(x)
\delta\chi(y)} \Big|_{\chi={\bar\chi} =0} \; .
\end{equation}
Considering only one-loop corrections given by (\ref{eq:gamma1.3}), we have
\begin{equation}
  \label{eq:quantumD2} 
{\tilde{\mathcal D}}(x,y)\;=\;{\mathcal D}(x,y) \;+\;g^2\,\Delta_\varphi (y,x)\,
{\mathcal D}^{-1}(x,y) \;.
\end{equation}
In Figure~\ref{feynman} we show the Feynman diagram that represents the first order quantum
corrections to the fermion self-energy (second term in 
(\ref{eq:quantumD2})).

Using the expansions
$$
\langle x|{\mathcal D}^{-1}|y\rangle\;=\; \int \frac{d^2{\hat
    p}}{(2\pi)^2} \,e^{i {\hat p}\cdot({\hat x}-{\hat y})} \left\{
  \langle x_2|\psi_0\rangle \langle\psi_0|y_2\rangle \frac{-i \not \!
    {\hat p}}{{\hat p}^2}{\mathcal P}_R \right.
$$
$$
+\sum_{n=1}^\infty [\langle
x_2|\psi_n\rangle\langle\psi_n|y_2\rangle \frac{-i \not \! {\hat
    p}}{{\hat p}^2+\lambda_n^2}{\mathcal P}_R +\,\langle x_2|{\tilde
  \psi}_n\rangle \langle{\tilde \psi}_n|y_2\rangle \frac{-i \not \!
  {\hat p}}{{\hat p}^2+\lambda_n^2} {\mathcal P}_L
$$
\begin{equation}
  \label{eq:ferexp}
\left. +\,\sum_{n=1}^\infty [\langle x_2|{\tilde \psi}_n\rangle
\langle\psi_n|y_2\rangle \frac{\lambda_n}{{\hat p}^2+\lambda_n^2} 
{\mathcal P}_R +\,\langle x_2|\psi_n\rangle \langle{\tilde \psi}_n|y_2\rangle 
\frac{\lambda_n}{{\hat p}^2+\lambda_n^2} 
{\mathcal P}_L ] \right\} 
\end{equation}
and
\begin{equation}
  \label{eq:bosexp}
\langle x|\Delta_\varphi|y\rangle \;=\; - \int \frac{d^2{\hat k}}{(2\pi)^2} \,
e^{i {\hat k}\cdot({\hat x}-{\hat y})}
\sum_{n=0}^\infty \langle x_2|\xi_n\rangle \langle\xi_n|y_2\rangle 
\frac{1}{{\hat k}^2 + \mu_n^2}\;,
\end{equation}
we obtain
\begin{equation}
  \label{eq:quantumD3}
{\tilde{\mathcal D}}(x,y)\;=\;{\mathcal D}(x,y)+\int\frac{d^2{\hat k}}{(2\pi)^2} \,
e^{i {\hat k} \cdot ({\hat x} - {\hat y})} \Upsilon_k (x_2,y_2) 
\end{equation}
with
$$
\Upsilon_k (x_2,y_2) \;=\; - g^2 \, \left\{
 \sum_{n=0}^\infty \,
 \xi_n(x_2) \psi_0(x_2) \xi^\dagger_n(y_2) \psi^\dagger_0(y_2) \,
  J({\hat k};\mu_n,0) {\mathcal P}_R \right.
$$
$$
+\,\sum_{n=0,m=1}^\infty\,[\xi_n(x_2)\psi_m(x_2)\xi^\dagger_n(y_2) 
\psi^\dagger_m(y_2) \, J({\hat k};\mu_n,\lambda_m) {\mathcal P}_R ] 
$$
$$
+\,\sum_{n=0, m=1}^\infty \, \xi_n(x_2) {\tilde\psi}_m(x_2) \xi^\dagger_n(y_2) 
{\tilde\psi}^\dagger_m(y_2) \, J({\hat k};\mu_n,\lambda_m) {\mathcal P}_L  
$$
$$
+ \, \sum_{n=0, m=1}^\infty \, \xi_n(x_2) {\tilde\psi}_m(x_2) \xi^\dagger_n(y_2) 
\psi^\dagger_m(y_2) \, I({\hat k};\mu_n,\lambda_m) {\mathcal P}_R  
$$
\begin{equation}
  \label{eq:gammak}
\left. + \, \sum_{n=0, m=1}^\infty \, \xi_n(x_2) \psi_m(x_2) \xi^\dagger_n(y_2) 
{\tilde\psi}^\dagger_m(y_2) \, I({\hat k};\mu_n,\lambda_m) {\mathcal P}_L \right\} \;,
\end{equation}
and the loop integrals
\begin{equation}
  \label{eq:defi}
I({\hat k}; M_1, M_2)\;=\; \int \frac{d^2 {\hat p}}{(2 \pi)^2}\,
\frac{M_2}{[({\hat k}-{\hat p})^2 + M_1^2]({\hat p}^2 + M_2^2)} 
\end{equation}

\begin{equation}
  \label{eq:defj}
J({\hat k}; M_1, M_2)\;=\; \int \frac{d^2 {\hat p}}{(2 \pi)^2}\,
\frac{-i\not \! {\hat p}}{[({\hat k}-{\hat p})^2 + M_1^2]({\hat p}^2 + M_2^2)} \;.
\end{equation}
We used a discrete notation for the sum over eigenvalues, but
of course an integral over the continuous part of the spectrum is
implicitly assumed.

At this stage we recognize, as was anticipated, that the Dirac operator
(\ref{eq:quantumD3}) is of the form given by (\ref{eq:defDirac2}) and
(\ref{eq:defnonlocalmass}).


\subsection{Quantum corrections to the fermionic zero
mode}\label{subsec:quantumcorr}

In the previous subsection we saw how the quantum fluctuations generate 
non localities in the Dirac operator (\ref{eq:quantumD3}); now we want to find its zero mode
$\Psi(x)$: $\langle x|\tilde{\mathcal D} | \Psi \rangle \,=\,0 , \;$ and compare it
with (\ref{eq:azeromode}). Repeating the steps that led to
(\ref{eq:nonlocala1}), we obtain $\lambda=g^2$ and
\begin{equation}
  \label{eq:gammak2}
\gamma_k(x_2,y_2)\,=\,\sum_{n=0, m=1}^\infty \, \xi_n(x_2) \, \psi_m(x_2) \,
\xi^\dagger_n(y_2) \,{\tilde\psi}^\dagger_m(y_2) \, I({\hat k};\mu_n,\lambda_m) \;,
\end{equation}
because in $\Upsilon_k$ only the term proportional to $P_L$ and $I({\hat
  k};\mu_n,\lambda_m)$ gives a non zero contribution to
(\ref{eq:defzeromode1}).  In effect, $P_R \chi \,=\,0$ and the term
proportional to $J(\hat k; \mu_n,\lambda_m) \, \propto \hat{\not \! k}$ also yields
a vanishing contribution when applied to $\chi$, after integration over
$x_2$.

On the other hand, for $n=0$, $I(\hat k; \mu_0,\lambda_m)\,=\,I(\hat
k; 0, \lambda_m)$ is IR divergent. This divergence can be associated with the
massless bosonic zero mode 
\begin{equation}
\xi_0 (x_2)\,=\,\sqrt{\frac{3 \kappa}{4}} \, \rm{sech}^2(\kappa x_2)
\end{equation}
living in a defect of infinite extension in the
$x_2$-direction; it contributes to the propagator $\Delta_{\varphi}$ in the
Feynman graph in Figure~\ref{feynman}.
Using an IR cutoff $\epsilon \to 0$, the divergent part of the regularized loop integral becomes
\begin{equation}
  \label{eq:Ireg}
I_{\epsilon}(\hat k;0,\lambda_m)=-\frac{1}{2 \pi} \, \frac{\lambda_m}{\hat k ^2 + \lambda_m ^2} \, \ln \epsilon \; .
\end{equation}
In other words, in the limit of infinite size, the dominant
contribution to $\gamma_k$ comes from the zero mode, and we can write
\begin{equation}
  \label{eq:gammanoncompact2}
\gamma_k(x_2,y_2) \approx \, \Big( \sum_{m=1}^{\infty} I_{\epsilon}(\hat k;0,\lambda_m)\, \psi_m (x_2) \, \tilde{\psi}_m ^{\dag} (y_2) \Big )  \, \xi_0 (x_2) \, \xi_0 ^{\dag}(y_2)
\,  \; .
\end{equation}

Furthermore, the $m>1$ modes give exponentially decreasing corrections in $\lambda_m$. This can be seen by applying the quantum Dirac operator $\tilde {\mathcal D}$ to the zero mode $\psi_0$ (obtained in Eq. (\ref{eq:zeromode3})) of the classical operator $\mathcal D$:
\begin{equation}
  \label{eq:correcexp}
\langle x_2|{\tilde {\mathcal D}}|\psi_0\rangle \,=\, 0 + \int_{-\infty}^{+\infty}dy_2\, \gamma_k(x_2,y_2) \langle y_2|\psi_0\rangle \;.
\end{equation}
From (\ref{eq:gammanoncompact2}), the integral over $y_2$ in (\ref{eq:correcexp}) is proportional to
\begin{equation}
\int_{-\infty}^{+\infty}dy_2\, {\tilde \psi}^\dagger_m (y_2) \psi_m(y_2) \xi^\dagger_0(y_2) \;=\; \,\frac{3 \sqrt{\pi}}{8} \, (\frac{\lambda_m}{\kappa})^2 \, 
{\rm sech}[\frac{\pi}{2} \sqrt{(\frac{\lambda_m}{\kappa})^2 -1}]  \;.
\end{equation}

Consequently, we can approximate (\ref{eq:gammanoncompact2}) by considering only the $m=1$ term:
\begin{equation}
  \label{eq:gammanoncompact3}
\gamma_k(x_2,y_2) \approx \, I_{\epsilon}(\hat k;0,\kappa)\, \xi_0 (x_2) \, \xi_0 ^{\dag}(y_2) \psi_1 (x_2) \, \tilde{\psi}_1 ^{\dag} (y_2) \, 
\,  \; .
\end{equation}
Later in this section, we will indicate how to construct the corrected zero mode when other terms in (\ref{eq:gammanoncompact2}) are taken into account. The presence of $\psi_1$ in $\gamma_k(x_2,y_2)$ tells us that quantum
fluctuations will mix the first two modes corresponding to the local mass case.
Substituting (\ref{eq:gammanoncompact3}) into (\ref{eq:kernel1}), we see that
\begin{equation}
  \label{eq:kernel2}
\tilde{\gamma}_k (x_2,y_2)\,=\,\frac{3 \kappa}{8 \pi} \, I_{\epsilon}(\hat k; 0, \kappa) \,
\big(1-\rm{sech}(\kappa x_2) \big) \rm{sech}^3 (\kappa y_2) \; .
\end{equation}

This is a separable kernel, because it is of the form $\tilde{\gamma}_k (x_2,y_2)\,=\,
\alpha(x_2) \, \beta(y_2)\,$. The Fredholm integral equation with separable
kernel may be solved, without using the Fredholm alternative~\cite{fredholm}, by
the following method.
Defining $C \equiv \frac{3 \kappa}{8 \pi} g^2 \, I_{\epsilon}(\hat k; 0, \kappa)$ we
obtain, from (\ref{eq:fredholm1})
\begin{equation}
  \label{eq:fredholm4}
\varphi(x_2)+C \big( {\rm{sech}}(\kappa x_2)-1\big) \, \int_{-\infty}^{+\infty} \, dy_2
\; {\rm{sech}}^3(\kappa y_2) \, \varphi(y_2)\,=\, \varphi(0) \;.
\end{equation}

Multiplying (\ref{eq:fredholm4}) by $\rm{sech}^3( \kappa x_2)$, integrating over
$x_2$, and introducing the constant
\begin{equation}
  \label{eq:defX}
X \equiv \int_{-\infty}^{+\infty} \, dy_2
\; {\rm sech}^3(\kappa y_2) \, \varphi(y_2)
\end{equation}
the result is
\begin{equation}
X \Big(1-\frac{C}{\kappa}(\frac{\pi}{2}-\frac{4}{3}) \Big)\,=\, \frac{\pi}{2
\kappa} \, \varphi(0) \; .
\end{equation}
As $C \gg 1$, 
\begin{equation}
  \label{eq:X2}
-C \, X \approx \frac{\pi}{\pi-8/3} \, \varphi(0) \; .
\end{equation}
On the other hand, from (\ref{eq:fredholm4}),
\begin{equation}
\varphi(x_2)+\big(1-\rm{sech}(\kappa x_2) \big) \, (-C \, X)\,=\, \varphi(0) \;.
\end{equation}
Therefore we arrive to a result which is independent of $C$ in the large size
limit:
\begin{equation}
  \label{eq:varphi}
\varphi(x_2) \approx \varphi(0) \Big[ 1+\frac{\pi}{\pi-8/3} \; \big({\rm sech}(\kappa
x_2) -1 \big) \Big] \;.
\end{equation}

The normalized zero mode including quantum corrections is, according to
(\ref{eq:zeromode3})
\begin{equation}
  \label{eq:psiquantum}
\psi(x_2) \,=\, N \Big[ 1+\frac{\pi}{\pi-8/3} \; \big( {\rm sech}(\kappa
x_2) -1 \big) \Big] \, {\rm sech}(\kappa x_2) \; ,
\end{equation}
where $N$ is a numerical normalization constant.

It should be clear from the previous derivation that the infrared
singularity that arises in the loop integral $I(\hat k; 0, \kappa)$ does
not propagate to the physical results; namely, those obtained for
large values of the infrared cutoff.

\begin{figure}
\begin{center}
\begin{picture}(0,0)%
\includegraphics{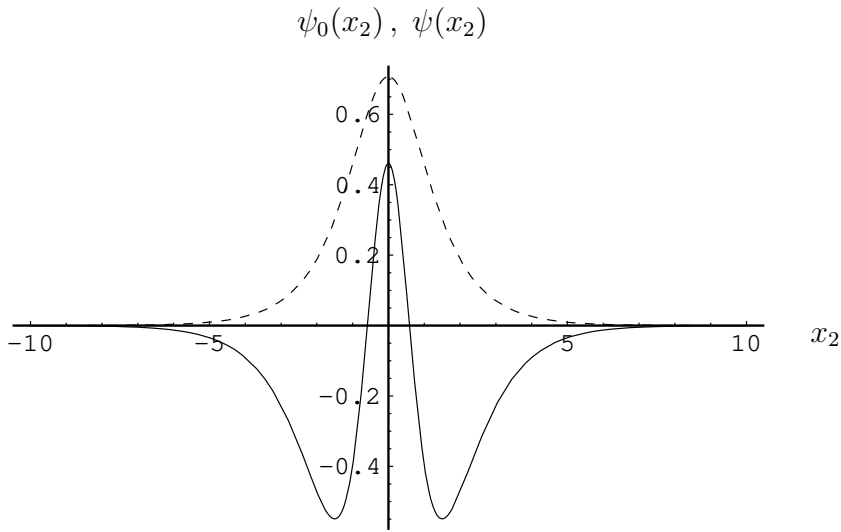}%
\end{picture}%
\setlength{\unitlength}{3947sp}%
\begingroup\makeatletter\ifx\SetFigFont\undefined%
\gdef\SetFigFont#1#2#3#4#5{%
  \reset@font\fontsize{#1}{#2pt}%
  \fontfamily{#3}\fontseries{#4}\fontshape{#5}%
  \selectfont}%
\fi\endgroup%
\begin{picture}(5048,3300)(1478,-12292)
\put(6526,-11086){\makebox(0,0)[lb]{\smash{\SetFigFont{12}{14.4}{\rmdefault}{\mddefault}{\updefault}{\color[rgb]{0,0,0}$x_2$}%
}}}
\put(3301,-9136){\makebox(0,0)[lb]{\smash{\SetFigFont{12}{14.4}{\rmdefault}{\mddefault}{\updefault}{\color[rgb]{0,0,0}$\psi_0 (x_2) \, , \; \psi(x_2)$}%
}}}
\end{picture}
\end{center}
\caption{\footnotesize{Classical (dashed line) and quantum-corrected
    (full line) zero mode profiles.}}
\label{modo2}
\end{figure}

In Figure~\ref{modo2} we plot $\psi_0 (x_2) \;$ and $\psi(x_2)\;$ (given by
(\ref{eq:psiquantum})); there, it becomes evident the quantum mixture of the
eigenmodes $\psi_0$ and $\psi_1$. In addition, the dispersions corresponding to
the distributions $|\psi_0(x_2)|^2$ and $|\psi(x_2)|^2$ are $\sigma_{loc} \approx
0.907/\kappa$ and $\sigma_{nonloc} \approx 1.850/\kappa \;$, respectively. Since
the local fermion mass term is $\kappa \, \rm{\rm{tanh}(\kappa x_2)}$, $\kappa$ is
the parameter that controls the localization of the domain wall and
consequently of the fermion zero modes, as seen from the previous dispersion
formulas. 

Besides, from Figure~\ref{modo2} it is clear that the quantum
correction given by (\ref{eq:psiquantum}) are nonperturbative; in
fact, this result cannot be obtained by resuming the Neumann series
(\ref{eq:neumann}). Quantum effects mix the zero localized mode $\psi_0$
with the nonlocalized mode $\psi_1$ giving, as a result, a nontrivial
localized function with two nodes and broader than the local zero mode
one. Of course, one would expect the inclusion of more modes should
result in an extra distorsion of the zero mode.

The generalization of this calculation to include a finite number of
terms from (\ref{eq:gammanoncompact2}), associated to the contribution
of the $\psi_{m>1}$ modes, should proceed as follows: applying the same
procedure as before to an expression of the form
\begin{equation}
\varphi(x_2)+ \sum_m a_m (x_2) \int_{-\infty}^{+\infty} dy_2 \, e_m (y_2) \, \varphi(y_2)\,=\,\varphi(0) \;,
\end{equation}
leads to the linear system
\begin{equation}
X_n+\sum_m a_{nm} X_m  \,=\, c_n \; ,
\end{equation}
where
$$
X_n \,=\,\int _{-\infty}^{+\infty} dy_2 \, e_n (y_2) \, \varphi(y_2) \; \;, \; a_{nm}=\int _{-\infty}^{+\infty} dy_2 \, e_n (y_2) \, a_m (y_2) \; ,
$$
\begin{equation}
c_n= \varphi(0) \int_{-\infty}^{+\infty} dy_2 \, e_n (y_2) \;.
\end{equation}
In terms of  the solutions $\{X_n\}$, $\varphi$ would become
\begin{equation}
\varphi(x_2)\,=\, \varphi(0)- \sum_m X_m \, a_m(x_2) \;.
\end{equation}

We conclude this section with a comment on the relation between 
${\rm dim} \, {\rm ker} (a)$ and
${\rm dim} \, {\rm ker} (a^{\dag})$. In the local case, only $a$ has a normalizable zero
mode; consequently the index relation
\begin{equation}
  \label{eq:index}
{\rm dim} \, {\rm ker} (a) - {\rm dim} \, {\rm ker}  (a^{\dag}) \, =\, 1
\end{equation}
is satisfied. This is characteristic of a harmonic oscillator. We saw
in (\ref{eq:psiquantum}) that the quantum-corrected destruction operator $a$ also
has a normalizable zero mode; besides, it's easy to see that the corrected
operator $a^{\dag}$ does not have such a mode. As a result, quantum fluctuations
to one-loop order preserve the relation (\ref{eq:index}). It should be noted
that in the compactified case, both $a$ and $a^{\dag}$ have zero modes, so
\begin{equation}
{\rm dim} \, {\rm ker} \, a - {\rm dim }\, {\rm ker} \, a^{\dag} \, =\, 0 \; .
\end{equation}
The consequences of these index relations are studied, for example, in~\cite{index}.

\section{Conclusions}\label{sec:conc}
We have analyzed a generalization of the Callan-Harvey mechanism to
the case of a non local mass in 2+1 dimensions, showing that for a
certain set of assumptions about the non locality, there continues to
exist a fermionic zero mode. 

We have first considered a quite general non local term, deriving a
linear integral equation for a function which modulates the zero mode
of the local case, and accounts for the effect of the non local domain
wall.  Considering a defect of finite size, the Fredholm alternative
theorem applies and there is a unique, localized chiral zero
mode. Moreover, perturbation theory may be applied to calculate it: 
for the concrete example of a strongly diagonal mass, we have calculated 
the first few terms in a perturbative expansion, showing that they lead to
negligible modifications with respect to the local mass case.

Finally, we have shown how radiative corrections for a system with a
Dirac field coupled to a scalar field do generate a non local mass in
the Dirac operator.  In this case, we have been able to derive non
perturbative quantum corrections to the fermionic zero mode, which
should be relevant in the infinite size limit.  That solution shows
the phenomenon of a `splitting' of the zero mode for the local case
into a corrected zero mode in three stripes, as seen in
Figure~\ref{modo2}. Of course, this is a reflection of the fact that 
the quantum corrected zero mode is a minimum of the {\em one loop 
effective action}, rather than of the classical action.

\section*{Aknowledgments}
G.~T.\ is supported by CNEA, Argentina. C.~D.~F.\ is supported by
CONICET (Argentina), and by a Fundaci\'on Antorchas grant.

\end{document}